\documentclass[notitlepage,a4paper,12pt]{article}
\usepackage{amsfonts}
\usepackage{amssymb}
\usepackage{eurosym}
\usepackage{harvard}
\usepackage{amsmath}
\usepackage{geometry}
\usepackage[onehalfspacing]{setspace}

\newtheorem{theorem}{Theorem}

\newtheorem{lemma}{Lemma}

\newenvironment{proof}[1][Proof]{\noindent \textbf{#1.} }{\  \rule{0.5em}{0.5em}}
\input{tcilatex}
\begin{document}

\title{On the Optimality of Full Disclosure\thanks{%
We thank Anton Kolotilin, Mikhail Drugov, and the participants of HSE\
Economic Theory workshop and EEA-ESEM 2022 congress, for helpful comments.}}
\author{Emiliano Catonini\thanks{%
NYU Shanghai. Email: emiliano.catonini@nyu.edu.} and Sergey Stepanov\thanks{%
HSE University, International College of Economics and Finance and Faculty
of Economic Sciences, Russia. Postal address: Office S1039, Pokrovsky
Boulevard 11, 109028 Moscow, Russia. Email: sstepanov@hse.ru}}
\maketitle

\begin{abstract}
A privately-informed sender can commit to any disclosure policy towards a
receiver. We show that full disclosure is optimal under a sufficient
condition with some desirable properties. First, it speaks directly to the
utility functions of the parties, as opposed to the indirect utility
function of the sender; this makes it easily interpretable and verifiable.
Second, it does not require the sender's payoff to be a function of the
posterior mean. Third, it is weaker than the known conditions for some
special cases. With this, we show that full disclosure is optimal under
modeling assumptions commonly used in principal-agent papers.


\end{abstract}

\section{Introduction}

We consider the classical problem of information transmission between a
sender with private, payoff-relevant information and a receiver who takes
actions which affect the sender's payoff. Following the Bayesian persuasion
literature pioneered by Rayo and Segal (2010) and Kamenica and Gentzkow
(2011), we suppose that the sender has commitment power over the information
she reveals to the receiver. Without setting any restrictions on the
possible persuasion strategies, we search for conditions under which full
disclosure is optimal. Differently from other complicated schemes, just
disclosing the truth seems to be a realistic goal in many scenarios ---
e.g., with transparency policies in organizations.

In our model, the state space can be a continuum, therefore the
concavification approach of Kamenica and Gentzkow (2011) is not operational.
Moreover, differently from most contributions in the field (e.g., Dworczak
and Martini (2019), Dizdar and Kov\'{a}\v{c} (2020), Gentzkow and Kamenica
(2016), Kolotilin et al. (2021), Arieli et al. (2020)), we do not assume
that the sender's payoff is a function of the posterior mean state (or any
moments of the posterior distribution).\footnote{%
We will discuss three notable exceptions in detail.} Despite this, we obtain
a sufficient condition for the optimality of full disclosure that speaks
directly to the underlying incentives of the parties, as opposed to the
indirect utility function of the sender. This makes our condition easily
interpretable and verifiable. In particular, it can be interpreted as a
requirement of minimal alignment of incentives between the sender and\ the
receiver. Notably, despite its level of generality, our condition is
substantially weaker than the sufficient condition provided by Kolotilin et
al. (2022) for environments in which the receiver's optimal action is linear
in the expected state.

\bigskip

To see why the effect of full disclosure may be non-trivial, consider a
simple principal-agent setup, as an example. The agent generates an output,
which he shares with the principal in a fixed proportion. The output is
increasing in the agent's effort, and the agent bears the cost of effort.
The state of nature determines the productivity of effort, with a higher
state resulting in higher productivity. The principal knows the state, while
the agent does not. At first sight, the principal would always want to
commit to revealing the state to the agent, as both parties \emph{seem} to
benefit from effort more when the state is higher. Here is a simple argument
why this may not be the case. Suppose that the agent is sufficiently risk
averse. Then, good news about the productivity may actually depress effort.
This is because a higher productivity implies that the agent reaches a
higher income, hence a lower marginal utility, at lower levels of effort. If
the principal is risk neutral, then the disclosure discourages the agent
precisely when the principal benefits more from effort (and incentivizes the
agent when the principal gains less from effort). In such a case, full
disclosure is unlikely to be optimal. Note also that, even when this
\textquotedblleft income effect\textquotedblright\ does not prevail in the
agent's incentives, full disclosure may still not be optimal. Even if the
agent increases effort under the good news and reduces it under the bad news
about the state, as the principal wants, the increase may be smaller than
the decrease, to the point that the overall effect on the principal's
utility\ is negative.

To see how we tackle these difficulties, stick to the principal-agent setup
and consider a message that pools two equally likely states. The principal
contemplates splitting this message into two messages that reveal the state.
Then, given the optimal effort under the pooling message, the agent will
discover that her marginal utility of effort is positive when one state is
revealed, negative when the other state is revealed, and the two values have
the same magnitude, just opposite signs. Thus, the agent will decrease
effort under the first state and increase it under the second state. Two
forces determine whether the principal gains from the split or not: the
changes in the agent's effort and the changes in the principal's utility per
unit of effort. Under each state, the agent modifies his effort until its
marginal utility returns to zero. Then what matters is how much the
principal's utility changes per unitary change of the agent's marginal
utility. In particular, the principal benefits from the split if this
measure of her marginal utility is larger when the agent wants to increase
effort with respect to when he prefers to reduce effort. In this sense, ours
is a condition of minimal alignment of interest between the two parties.

Our main result extends this argument to all possible messages in a general
sender-receiver framework. Specifically, we show that \emph{any} message
with a non-singleton support can be split so as to improve the sender's
welfare if an increase of action that decreases the receiver's marginal
utility by one unit has a larger benefit for the sender when it also
benefits the receiver, compared to when it harms him. This condition ensures
the optimality of full disclosure.

Under some additional regularity assumptions, we also provide an analogous
sufficient condition that is entirely expressed in terms of derivatives of
the parties' utility functions (\textquotedblleft derivatives
condition\textquotedblright ). This condition may be easier to check in some
economic applications.

Finally, we also derive a sufficient condition for the \emph{sub}optimality
of full disclosure. While there remains a gap\ between this condition and
our optimality condition (one is not a negation of the other), it helps to
establish when full disclosure is definitely not optimal, as we will show in
an example.

We then focus on the principal-agent setting we outlined before. Typically,
in this application, the principal's utility cannot be represented as a
function of the posterior mean. We discuss several examples demonstrating
that our sufficient condition for full disclosure is easy to check and often
satisfied. The first example (section \ref{CRRA}) sheds light on the role of
risk aversion for the optimality/suboptimality of full disclosure. We assume
that both parties exhibit CRRA and the output is a product of the state and
a concave power function of effort. Full disclosure turns out to be optimal
when the agent is more risk averse than the principal (a typical textbook
situation) but not too risk averse (with the coefficient of relative risk
aversion\ below one). In this case, state and effort are complements for
both parties, and then disclosing the state boosts effort exactly when the
principal benefits from higher effort more. Instead, when the agent becomes
too risk averse (while the principal remains moderately risk averse), full
disclosure ceases to be optimal. As we discussed earlier, under high agent's
risk aversion, good news about productivity \emph{depress} effort, that is,
effort and state become substitutes for the agent while remaining
complements for the principal.

Another interesting case discussed in Section \ref{CRRA} is when the agent
is sufficiently risk averse, and the principal is at least as risk averse as
the agent. In that case, the average effort falls but the principal
nevertheless gains from transparency. This happens because for the principal
effort and state are even more substitutes than for the agent. Bad news
about productivity encourages effort, and the principal benefits even more
from effort in lower states than the agent does.

In the second example (section \ref{risk neutral separable}) we simplify the
preferences by assuming risk neutrality for both parties and focus instead
on the properties of the production function that ensure the optimality of
full disclosure. By applying the \textquotedblleft derivatives
condition\textquotedblright , we show that full disclosure is optimal under
some commonly used functional forms for output.

\bigskip

Without assuming that the sender's payoff is a function of the expected
state, Kolotilin (2018) and Kolotilin et al. (2022) establish that (under
some assumptions on the utility functions) full disclosure is optimal if and
only if, for any pair of states, the sender prefers revealing them to
garbling.\footnote{%
In more rigorous terms, the sender prefers to split any posterior with a
binary support into two degenerate posteriors.} So, under some conditions,
the problem reduces to checking only messages with binary support. In some
simple cases (for example, the receiver's optimal action depends only on the
expected state and the sender's direct utility depends only on the action),
the sender's indirect utility function becomes a function of only the
posterior mean; then, the necessary and sufficient condition for the
optimality of full disclosure boils down to requiring the convexity of this
function. Kolotilin et al. (2022) make further progress by providing
sufficient conditions for the optimality of full disclosure on the sender's
(direct)\ utility function for the special case in which the receiver's
optimal action is linear in the expected state (while the sender's utility
is allowed to depend on the state as well as the action).

Except for the requirement that the receiver's utility is strictly concave
in action and delivers an interior solution (along with some regularity
assumptions), we impose no restrictions on how state and action affect
utilities. Despite this, we offer a sufficient condition for the optimality
of full disclosure in terms of the primitives of the model: the (direct)
utility functions of the sender and the receiver. In this way, compared to
the condition in Kolotilin (2018) and Kolotilin et al. (2022), we gain
operability and interpretability. Moreover, in contrast to Kolotilin (2018)
and Kolotilin et al. (2022), we do not impose a single-crossing assumption
on the receiver's utility. This allows applying our condition to
environments where considering only binary support messages may not be
without loss of generality. Despite this level of generality, our condition
turns out to be substantially weaker than the sufficient condition of
Kolotilin et al. (2022) for the case in which the receiver's action is
linear in the expected case, as it requires neither convexity of the
sender's payoff in action, nor its supermodularity in action and state.

\bigskip

Using a concept analogous to the concept of \textquotedblleft virtual
value\textquotedblright\ in the mechanism design literature, Mensch (2021)
offers conditions for full disclosure jointly on the receiver's utility
function and on a transformation of the sender's utility function that takes
into account the incentive compatibility constraint of the receiver
(\textquotedblleft virtual utility\textquotedblright ). His focus is on the
importance of complementarities between states and actions, and whether
these complementarities \textquotedblleft point in the same
direction\textquotedblright\ for the sender and the receiver. While Mensch's
condition for full disclosure (Theorem 5) is insightful, it is rather
abstract and not straightforward to apply, as it requires a derivation of
the \textquotedblleft virtual utility\textquotedblright . Instead, our
conditions are directly on primitives of the model, that is, the shape of
the parties' utility functions.

\bigskip

The paper is organized as follows. Section \ref{general model} sets up the
model. Section \ref{sufficient conditions} derives the conditions for the
optimality of full disclosure, as well as the condition for its
suboptimality. In Section \ref{simple cases}, we compare our sufficient
condition with the conditions obtained in the literature for two special
cases of the parties' preferences. Section \ref{PA model} demonstrates how
our conditions can be applied in a principal-agent setting and discusses the
role of risk aversion and complementarity/substitutability between the
action and the state. All proofs are relegated to the Appendix.

\section{Model\label{general model}}

There are a sender (she) and a receiver (he). The receiver needs to take a
non-contractible action $a\in A$. There is a state of the world $\omega \in
\Omega $ with common prior $p\in \Delta (\Omega )$. We assume that $A$ and $%
\Omega $ are compact intervals in the real line; with this, we do not rule
out that the possible states be discrete, because we do not impose
restrictions on the support of $p$.

Action and state jointly determine the receiver's utility $U(\omega ,a)$.%
\footnote{%
The state may just represent whatever information is available to the sender
about a \textquotedblleft more primitive\textquotedblright\ state that
affects payoffs; in this case $U$ is an expected utility.} We assume that,
for every $\omega \in \Omega $, $U(\omega ,a)$ is twice differentiable and
strictly concave in $a$, with $U_{a}(\omega ,a)=0$ for some finite $a\in A$,
denoted by $a^{\ast }(\omega )$. We also assume that $U_{a}(\omega ,a)$ is
continuous in $\omega $.

The sender's utility is $V(\omega ,a)$, and we assume it to be
differentiable in $a$, with $V_{a}(a,\omega )$ continuous in $\omega $.
Until Section \ref{PA model}, we abstract away from the origin of $U(\omega
,a)$ and $V(\omega ,a)$.

Before learning the state, the sender can commit to an information
structure, whereby the receiver gets some information about the state before
choosing the action. Formally, following the standard Bayesian persuasion
framework, the sender commits to a mapping from the set of states $\Omega $
to distributions over messages that are sent to the receiver. The
information structure chosen by the sender is common knowledge. The goal of
the sender is to select an information structure that maximizes her expected
utility.

\bigskip

After receiving message $m$, the receiver solves%
\begin{equation*}
\max_{a}\mathbb{E}(U(\omega ,a)|m)
\end{equation*}%
Due to our assumptions on $U(\omega ,a)$, the receiver's optimal action is
unique under every posterior belief about the state, and it is determined by
the first-order condition%
\begin{equation*}
\frac{d\mathbb{E}(U(\omega ,a)|m)}{da}=0
\end{equation*}%
By continuity of $U_{a}(\omega ,a)$ in $\omega $, the receiver's optimal
action changes continuously in the posterior belief. With this, the
persuasion problem of the sender is well-defined and has a solution.

\section{Sufficient conditions\label{sufficient conditions}}

\subsection{Main condition\label{main condition}}

Full disclosure is optimal if any message that pools or partially pools
several states that induce different actions can be split into several more
informative (in Blackwell sense) messages in a way that strictly increases
the sender's expected utility (conditional on the original message). We will
first consider messages that generate a posterior with binary support. The
crucial and most insightful passage of our construction identifies a
condition under which splitting a message with binary support into two
messages that reveal the state weakly benefits the sender --- we illustrate
this passage in detail in the main text (and report a more formal proof in
the Appendix). Then, we will sketch how we extend this argument to find a 
\emph{strictly} profitable split of \emph{any} message $m$ under a slightly
stronger condition, and finally establish the optimality of full disclosure
under the original condition --- the details of these two passages are
deferred to the formal proof in the Appendix.

Consider two states, $\omega _{1}$ and $\omega _{2}$, such that the
receiver's optimal action is higher under $\omega _{2}$: $a_{2}^{\ast
}:=a^{\ast }(\omega _{2})>a^{\ast }(\omega _{1})=:a_{1}^{\ast }$. Let $m$ be
a message that (partially) pools $\omega _{1}$ and $\omega _{2}$, and let $%
\pi _{1}:=\Pr (\omega _{1}|m),\ \pi _{2}:=\Pr (\omega _{2}|m),\ \pi
_{2}=1-\pi _{1}$. The graph below depicts the receiver's utilities under $%
\omega _{1},\ \omega _{2}$, and his expected utility under $m$: $U(\omega
_{1},a),\ U(\omega _{2},a),\ \widetilde{U}(\omega ,a)$. Action $a^{\ast }$
denotes the receiver's optimal action under $m$. The sender's
state-contingent utilities $V(\omega _{1},a)$ and$\ V(\omega _{2},a)$ are
depicted increasing, with $V(\omega _{2},a)$ above $V(\omega _{1},a)$, for
illustration purposes, but they do not have to be such.

\FRAME{dtbpFU}{5.4967in}{3.416in}{0pt}{\Qcb{Figure 1.}}{}{Figure 1}{\special%
{language "Scientific Word";type "GRAPHIC";maintain-aspect-ratio
TRUE;display "USEDEF";valid_file "T";width 5.4967in;height 3.416in;depth
0pt;original-width 6.4965in;original-height 4.0266in;cropleft "0";croptop
"1";cropright "1";cropbottom "0";tempfilename 'fig1.wmf';tempfile-properties
"XNPR";}}

Conditional on $m$, the sender (weakly) benefits from disclosing $\omega
_{1},\ \omega _{2}$ instead of sending $m$ if and only if%
\begin{equation*}
\pi _{1}V(\omega _{1},a_{1}^{\ast })+\pi _{2}V(\omega _{2},a_{2}^{\ast
})\geq \pi _{1}V(\omega _{1},a^{\ast })+\pi _{2}V(\omega _{2},a^{\ast }),
\end{equation*}%
that is,%
\begin{equation}
\pi _{2}[V(\omega _{2},a_{2}^{\ast })-V(\omega _{2},a^{\ast })]\geq \pi
_{1}[V(\omega _{1},a^{\ast })-V(\omega _{1},a_{1}^{\ast })].  \label{jumps}
\end{equation}%
Graphically, condition (\ref{jumps}) means that the probability-weighted
increase in the sender's payoff as we move from $A$ to $B$ exceeds the
probability-weighted decrease as we move from $C$ to $D$.\footnote{%
We are saying \textquotedblleft increase\textquotedblright\ and
\textquotedblleft decrease\textquotedblright\ to relate to the graph. But,
as we have said, $V(\omega ,a)$ does not have to be upward sloping, so, in
general, it is more accurate to talk about comparing the change as we move
from $A$ to $B$ with a negative of the change as we move from $C$ to $D$,
exactly as in (\ref{jumps}).}

If (\ref{jumps}) holds for all possible $\omega _{1},\ \omega _{2}$ and $\pi
_{1}$, full disclosure is optimal. Stated in this way, the condition does
not help much, as it does not provide a recipe to verify it for all possible 
$\omega _{1},\ \omega _{2}$ and $\pi _{1}$.

Our idea is as follows. First, instead of comparing the total
probability-weighted changes in the sender's state-contingent payoff, we are
going to compare \textquotedblleft marginal changes\textquotedblright\
(weighted with the corresponding probabilities) as we move from $A$ to $B$
and from $C$ to $D$, \textquotedblleft pointwise\textquotedblright .\ We
will define what it means for a change to be \textquotedblleft
marginal\textquotedblright\ in such a way that if any marginal change\ on
the way from $A$ to $B$ is larger than on the way from $C$ to $D$, the total
change will be larger as well.

Second, notice that any $a$ on the way from $A$ to $B$ (i.e., between $%
a^{\ast }$ and $a_{2}^{\ast }$), is higher than any $a$ on the way from $C$
to $D$ (i.e., between $a^{\ast }$ and $a_{1}^{\ast }$). In addition, $%
U_{a}(\omega _{2},a)>0$ for any $a\in \lbrack a^{\ast },a_{2}^{\ast })$, and 
$U_{a}(\omega _{1},a)<0$ for any $a\in (a_{1}^{\ast },a^{\ast }]$. Since
these properties hold for any message with binary support, they allow us to
formulate a sufficient condition that neither involves specific posterior
probabilities nor requires computing the optimal receiver's action.

We start from defining the marginal changes. We cannot compare marginal
changes in the space of $a$, because $[a_{1}^{\ast },a^{\ast }]$ and $%
[a^{\ast },a_{2}^{\ast }]$ have different lengths. Hence, we move to the
space of probability-weighted receiver's marginal utilities: $x_{1}:=\pi
_{1}U_{a}(\omega _{1},a)$ and $x_{2}:=-\pi _{2}U_{a}(\omega _{2},a)$. As $a$
runs from $a^{\ast }$ to $a_{1}^{\ast }$ (for $x_{1}$) and from $a^{\ast }$
to $a_{2}^{\ast }$ (for $x_{2}$), both $x_{1}$ and $x_{2}$ run from the 
\emph{same} constant, $k<0$, to zero. That the starting point is the same
stems from the first-order condition under $m$:%
\begin{eqnarray}
\pi _{1}U_{a}(\omega _{1},a^{\ast })+\pi _{2}U_{a}(\omega _{2},a^{\ast })
&=&0  \label{FOC under pooling} \\
&\Rightarrow &\pi _{1}U_{a}(\omega _{1},a^{\ast })=-\pi _{2}U_{a}(\omega
_{2},a^{\ast })=:k  \notag
\end{eqnarray}%
That the arrival point is zero is due to the first-order condition\ under $%
\omega _{i}$: $U_{a}(\omega _{i},a_{i}^{\ast })=0$.

Now, since $x_{1}$ and $x_{2}$ span the same intervals, comparing marginal
changes in $V(\omega _{1},a)$ and $V(\omega _{2},a)$ in the space of $x_{1}$
and $x_{2}$ (respectively) is legitimate. Comparing a marginal gain from
revealing $\omega _{2}$ with a marginal loss from revealing $\omega _{1}$%
\footnote{%
Here again it would be more accurate to say \textquotedblleft a marginal
change\textquotedblright\ from revealing $\omega _{2}$ and \textquotedblleft
negative of a marginal change\textquotedblright\ from revealing $\omega _{1}$%
.} at given $a_{1}\in (a_{1}^{\ast },a^{\ast })\ $and $a_{2}\in (a^{\ast
},a_{2}^{\ast })$, is the same as comparing $\partial (\pi _{2}V(\omega
_{2},a_{2}))/\partial x_{2}(a_{2})$ with $-\partial (\pi _{1}V(\omega
_{1},a_{1}))/\partial x_{1}(a_{1})$ (we are using $\partial $ to emphasize
that we are differentiating while holding $\omega _{i}$ and $\pi _{i}$
fixed). Thus, if%
\begin{gather*}
\partial (\pi _{2}V(\omega _{2},a_{2}))/\partial x_{2}(a_{2})\geq -\partial
(\pi _{1}V(\omega _{1},a_{1}))/\partial x_{1}(a_{1}) \\
\text{for all }a_{1}\in (a_{1}^{\ast },a^{\ast })\text{ and}\ a_{2}\in
(a^{\ast },a_{2}^{\ast }),
\end{gather*}%
inequality (\ref{jumps}) will be satisfied.

Now, notice that\footnote{%
Formally, the second equality in each of the two lines below can be derived
as follows. Let $y=U_{a}(\omega ,a)$ and $a=U_{a}^{-1}(\omega ,y)$
respectively. Then, holding $\omega $ fixed:%
\begin{equation*}
\frac{\partial V(\omega ,U_{a}^{-1}(\omega ,y))}{\partial y}=V_{a}(\omega
,U_{a}^{-1}(\omega ,y))\frac{\partial U_{a}^{-1}(\omega ,y)}{\partial y}=%
\frac{V_{a}(\omega ,U_{a}^{-1}(\omega ,y))}{U_{aa}(\omega ,U_{a}^{-1}(\omega
,y))}=\frac{V_{a}(\omega ,a)}{U_{aa}(\omega ,a)}.
\end{equation*}%
}%
\begin{eqnarray*}
\frac{\partial (\pi _{1}V(\omega _{1},a))}{\partial x_{1}(a)} &=&\frac{%
\partial V(\omega _{1},a)}{\partial U_{a}(\omega _{1},a)}=\frac{V_{a}(\omega
_{1},a)}{U_{aa}(\omega _{1},a)}, \\
\frac{\partial (\pi _{2}V(\omega _{2},a))}{\partial x_{2}(a)} &=&\frac{%
\partial V(\omega _{2},a)}{-\partial U_{a}(\omega _{2},a)}=\frac{%
V_{a}(\omega _{2},a)}{-U_{aa}(\omega _{2},a)}.
\end{eqnarray*}%
Moreover, notice that for all $a_{1}\in (a_{1}^{\ast },a^{\ast })$, $%
U_{a}(\omega _{1},a_{1})<0$, and for all $a_{2}\in (a^{\ast },a_{2}^{\ast })$%
, $U_{a}(\omega _{2},a_{2})>0$.

Consequently, if $-V_{a}(\omega _{2},a_{2})/U_{aa}(\omega _{2},a_{2})\geq
-V_{a}(\omega _{1},a_{1})/U_{aa}(\omega _{1},a_{1})$ for any $a_{1},\
a_{2},\ \omega _{1},\ \omega _{2}$ such that $a_{1}<a_{2}$ and $U_{a}(\omega
_{1},a_{1})<0<U_{a}(\omega _{2},a_{2})$, revealing the states in the support
of any binary-support message benefits the sender. Hence, we arrive at the
following sufficient condition for the optimality of splitting any message
with binary support:%
\begin{gather}
\text{For all }a_{1},\ a_{2},\ \omega _{1},\ \omega _{2}\text{,}  \notag \\
\hspace{0in}\left\{ 
\begin{array}{c}
a_{1}<a_{2} \\ 
\hspace{0in}U_{a}(\omega _{1},a_{1})<0<U_{a}(\omega _{2},a_{2})%
\end{array}%
\right. \Rightarrow \frac{V_{a}(\omega _{1},a_{1})}{-U_{aa}(\omega
_{1},a_{1})}\leq \frac{V_{a}(\omega _{2},a_{2})}{-U_{aa}(\omega _{2},a_{2})}.
\label{THE WEAK CONDITION}
\end{gather}%
Condition (\ref{THE WEAK CONDITION}) can be concisely phrased as the
requirement that $V_{a}(\omega ,a)/(-U_{aa}(\omega ,a))$ goes up (or stays
the same)\ whenever both $a$ and $U_{a}(\omega ,a)$ increase and $%
U_{a}(\omega ,a)$ switches from negative to positive.

\begin{lemma}
\label{binary split}Under condition (\ref{THE WEAK CONDITION}), for any
message that generates a posterior with binary support, revealing the states
in the support instead of sending the message weakly increases the expected
utility of the sender, conditional on the message. If the inequality between
the ratios in (\ref{THE WEAK CONDITION}) is strict, the expected utility of
the sender strictly increases.
\end{lemma}

Kolotilin (2018) shows that, under certain assumptions, it is enough to
consider only binary support messages to check for the optimality of full
disclosure.\footnote{%
Kolotilin (2018), Proposition 1, part (ii) and Corollary 1, part (ii). See
also Kolotilin et al. (2022), Lemma 3, for a more explicit formulation. More
precisely, both papers state that, under the assumptions that allow to focus
on binary-support messages, full disclosure is optimal if and only if (\ref%
{jumps}) holds for all possible $\omega _{1},\ \omega _{2}$ and $\pi _{1}$.
(By employing (\ref{FOC under pooling}), Kolotilin (2018) expresses the
condition in terms of $U_{a}(\omega _{1},a^{\ast })$ and $U_{a}(\omega
_{1},a^{\ast })$ instead of $\pi _{1}$ and $\pi _{2}$.) As we argued in the
Introduction, compared to these papers, our contribution consists of
translating the necessary-and-sufficient but abstract condition (\ref{jumps}%
) into a just sufficient but easily interpretable/verifiable condition, and
extending it to settings where considering binary-support messages may not
be enough.} These assumptions are: (i) both $A$ and $\Omega $ are compact
intervals in $%
\mathbb{R}
$, (ii) $U_{a}(\omega ,a)$ and $V_{a}(\omega ,a)$ are continuous in $\omega $
and continuously differentiable in $a$, (iii) for any posterior, the
receiver's expected utility is single-peaked in $a$ and his optimal $a$ is
interior, (iv) the receiver's optimal state-contingent action $a^{\ast
}(\omega )$ is monotonic in $\omega $ (\textquotedblleft single
crossing\textquotedblright ).

We have milder requirements compared to (ii), and, more importantly, our
framework does not impose (iv). So, we cannot rule out a priori that
non-binary support messages be unneeded to optimize the sender's utility.
Nonetheless, we are able to show that (\ref{THE WEAK CONDITION}) is a
sufficient condition for full disclosure, in the following way. First, we
extend the argument of Lemma \ref{binary split} to find a profitable split
of any arbitrary message $m$. To start, we show that we can always split $m$
into a message with binary support and a \textquotedblleft
complementary\textquotedblright\ message that both induce the same action as 
$m$. Then, if (\ref{THE WEAK CONDITION}) holds as a strict inequality, it is
tempting to say that a further split of the binary support message does the
job and generates a welfare-improving ultimate split. However, with a
continuous state space, the binary-support message may have a zero
probability conditional on $m$, and then we cannot claim welfare
improvement. We circumvent this problem by looking at arbitrarily small
\textquotedblleft neighborhoods\textquotedblright\ of the two states of the
binary-support message. This allows us to claim that (\ref{THE WEAK
CONDITION}) with the \emph{strict} instead of weak inequality is a
sufficient condition for the optimality of full disclosure. The last step
uses perturbations of the sender's utility function to claim that condition (%
\ref{THE WEAK CONDITION}) is sufficient for the optimality of full
disclosure. These steps are formalized in the proof of our main result:

\begin{theorem}
\label{main theorem 2}Under condition (\ref{THE WEAK CONDITION}) full
disclosure is optimal for the sender.
\end{theorem}

\bigskip

Condition (\ref{THE WEAK CONDITION}) does not require computing the
receiver's optimal response to a posterior and can be applied to a broad
class of sender's and receiver's utility functions (Section \ref{PA model}
provides examples). Moreover, it can be interpreted as a requirement of
minimal alignment of interest between the sender and the receiver. Suppose
for a second that $U_{aa}$ is a constant. Conditions $a_{1}<a_{2}$ and $%
U_{a}(\omega _{1},a_{1})<0<U_{a}(\omega _{2},a_{2})$ mean that state $\omega
_{2}$ generates positive incentives for the receiver (i.e., the incentive to
increase $a_{2}$) and state $\omega _{1}$ generates negative incentives
(i.e., the incentive to decrease $a_{1}$). Then, (\ref{THE WEAK CONDITION})
requires that the sender's marginal benefit from an increase in action is
(weakly) larger when such an increase is desirable for the receiver with
respect to when it is not.

\textquotedblleft Normalization\textquotedblright\ of $V_{a}$ by $U_{aa}$ in
(\ref{THE WEAK CONDITION}) can be understood as follows. It is important not
only how strong the sender's utility reacts to marginal changes in action,
but also how far the action moves once the state is revealed. The
\textquotedblleft speed of readjustment\textquotedblright\ is determined
precisely by $U_{aa}$. When $-U_{aa}(\omega _{2},a_{2})$ is lower, $a_{2}$
increases slower, that is, it goes a longer way until it reaches the optimal
value under $\omega _{2}$. This implies a higher benefit for the sender from
the revelation of $\omega _{2}$ if $V_{a}(\omega _{2},a_{2})$ is positive (a
higher loss if $V_{a}(\omega _{2},a_{2})$ is negative). Similarly, when $%
-U_{aa}(\omega _{1},a_{1})$ is lower, $a_{1}$ goes a longer way, but now
this is a \emph{decrease} towards the new optimal action, so there is a
higher loss from the revelation of $\omega _{1}$ if $V_{a}(\omega
_{1},a_{1}) $ is positive (a higher benefit if $V_{a}(\omega _{2},a_{2})$ is
negative).

Note also that condition (\ref{THE WEAK CONDITION}) is always trivially
satisfied when $V=U$, that is, when the incentives of the parties are
perfectly aligned. This is because $U_{a}(\omega _{1},a_{1})<0<U_{a}(\omega
_{2},a_{2})$ implies%
\begin{equation*}
\frac{U_{a}(\omega _{1},a_{1})}{-U_{aa}(\omega _{1},a_{1})}<\frac{%
U_{a}(\omega _{2},a_{2})}{-U_{aa}(\omega _{2},a_{2})},
\end{equation*}%
given that $U_{aa}<0$.

\subsection{Derivatives condition}

A stronger but somewhat simpler condition than (\ref{THE WEAK CONDITION}) is
the following:%
\begin{gather}
\text{For all }a_{1},\ a_{2},\ \omega _{1},\ \omega _{2}\text{,}  \notag \\
\left\{ 
\begin{array}{c}
a_{1}<a_{2} \\ 
U_{a}(\omega _{1},a_{1})<U_{a}(\omega _{2},a_{2})%
\end{array}%
\right. \Rightarrow \frac{V_{a}(\omega _{1},a_{1})}{-U_{aa}(\omega
_{1},a_{1})}\leq \frac{V_{a}(\omega _{2},a_{2})}{-U_{aa}(\omega _{2},a_{2})}.
\label{THE DERIVABLE CONDITION}
\end{gather}%
It is stronger than (\ref{THE WEAK CONDITION}) because it requires that the
relation between the ratios holds for a larger set of $(\omega
_{1},a_{1}),(\omega _{2},a_{2})$ pairs, where $U_{a}(\omega _{1},a_{1})$
does not have to be negative and $U_{a}(\omega _{2},a_{2})$ does not have to
be positive.

Assuming that $U_{aaa},\ U_{aa\omega }\ $and $V_{aa}$ exist, condition (\ref%
{THE DERIVABLE CONDITION}) can be expressed in terms of just derivatives of $%
U$ and $V$. To see this, notice that (\ref{THE DERIVABLE CONDITION}) is
equivalent to stating that, at each $(\omega ,a)$, $-V_{a}(\omega
,a)/U_{aa}(\omega ,a)$ is increasing in all directions in which both $a$ and 
$U_{a}(\omega ,a)$ increase. So, by applying directional derivatives, one
can show the lemma below. Namely, consider the following conditions:%
\begin{equation}
\text{For each }(\omega ,a)\text{ s.t. }U_{a\omega }>0\text{, }\left\{ 
\begin{array}{c}
U_{aa\omega }V_{a}\geq V_{a\omega }U_{aa} \\ 
V_{a}(U_{aaa}U_{a\omega }-U_{aa\omega }U_{aa})\geq U_{aa}(V_{aa}U_{a\omega
}-V_{a\omega }U_{aa})%
\end{array}%
\right. ,  \label{derivative conditions 1}
\end{equation}%
and%
\begin{equation}
\text{For each }(\omega ,a)\text{ s.t. }U_{a\omega }<0\text{, }\left\{ 
\begin{array}{c}
U_{aa\omega }V_{a}\leq V_{a\omega }U_{aa} \\ 
V_{a}(U_{aaa}U_{a\omega }-U_{aa\omega }U_{aa})\leq U_{aa}(V_{aa}U_{a\omega
}-V_{a\omega }U_{aa})%
\end{array}%
\right. .  \label{derivative conditions 2}
\end{equation}

\begin{lemma}
\label{derivative lemma}Assume that $U_{aaa},\ U_{aa\omega }$ and$\ V_{aa}$
exist. Then condition (\ref{THE DERIVABLE CONDITION}) is equivalent to (\ref%
{derivative conditions 1}) and (\ref{derivative conditions 2}).
\end{lemma}

Notice that (\ref{derivative conditions 1}) and (\ref{derivative conditions
2}) do not cover the case $U_{a\omega }=0$. This is because, when $%
U_{a\omega }=0$, there is simply no direction in which both $a$ and $U_{a}$
increase.

Subsection \ref{risk neutral separable} will illustrate the application of
the derivatives conditions.

\subsection{Sufficient condition for suboptimality of full disclosure}

Subsection \ref{main condition} delivered a sufficient condition for the
optimality of full disclosure. We can apply almost the same scheme of
reasoning to derive a sufficient condition for the \emph{sub}optimality of
full disclosure. Instead of the existence of a welfare-improving split for
any message with binary support, the suboptimality of full disclosure
requires the existence of at least one pair of states that can be pooled (or
partially pooled) so as to improve the sender's welfare.

Namely, fix a pair of states $\omega _{1}$, $\omega _{2}$ and consider the
following condition

\begin{equation}
\text{For all }a_{1},\ a_{2},\ \left\{ 
\begin{array}{c}
a_{1}<a_{2} \\ 
U_{a}(\omega _{1},a_{1})<0<U_{a}(\omega _{2},a_{2})%
\end{array}%
\right. \Rightarrow \frac{V_{a}(\omega _{1},a_{1})}{-U_{aa}(\omega
_{1},a_{1})}>\frac{V_{a}(\omega _{2},a_{2})}{-U_{aa}(\omega _{2},a_{2})}
\label{non-transparency}
\end{equation}%
This condition resembles (\ref{THE WEAK CONDITION}) except that it is
formulated for given $\omega _{1}$ and $\omega _{2}$ and the sign of the
inequality between the ratios flips.

\begin{theorem}
\label{non-transparency theorem}If there exists a pair of states $\omega
_{1},\omega _{2}\in \mathrm{supp}p$ such that (\ref{non-transparency}) holds
non-vacuously, full disclosure is suboptimal for the sender.
\end{theorem}

Notice that Theorem \ref{non-transparency theorem} does not imply that (\ref%
{THE WEAK CONDITION}) delivers a necessary and sufficient condition for the
optimality of full disclosure. The fact that $-V_{a}(\omega
_{1},a_{1})/U_{aa}(\omega _{1},a_{1})\leq -V_{a}(\omega
_{2},a_{2})/U_{aa}(\omega _{2},a_{2})$ fails to hold for some $a_{1},\
a_{2},\ \omega _{1},\ \omega _{2}$ such that $a_{1}<a_{2}$ and $U_{a}(\omega
_{1},a_{1})<0<U_{a}(\omega _{2},a_{2})$ does not mean that there will
necessarily be a pair of states $\omega _{1}$ and $\omega _{2}$ for which $%
-V_{a}(\omega _{1},a_{1})/U_{aa}(\omega _{1},a_{1})>-V_{a}(\omega
_{2},a_{2})/U_{aa}(\omega _{2},a_{2})$ for \emph{all} $a_{1},\ a_{2},\ $such
that $a_{1}<a_{2}$ and $U_{a}(\omega _{1},a_{1})<0<U_{a}(\omega _{2},a_{2})$%
, as the relation between the ratios may change sign as $a_{1}$ and $a_{2}$
change.

\section{Well-known special\ cases\label{simple cases}}

In this section we compare our sufficient condition with the conditions
derived in the literature for two specific cases.

\subsection{\textquotedblleft Linear case\textquotedblright}

Much of the literature has focused on settings in which the sender's payoff
from sending a certain message can ultimately be represented as a function
of the posterior mean only. This is the case, for example, when the
receiver's action only depends on the expected state, $\mathbb{E}(\omega |m)$%
, and the sender's utility only depends on the receiver's action: $V(\omega
,a)=V(a)$. Then, given the posterior induced by message $m$, the sender's
payoff is $V(a^{\ast }(\mathbb{E}(\omega |m))$, which can be represented as
an indirect utility function, $\widehat{V}(\mathbb{E}(\omega |m))$. It is
well known that the necessary and sufficient condition for the optimality of
full disclosure in this case is that $\widehat{V}(\cdot )$ is convex on the
set of admissible values for $\mathbb{E}(\omega |m)$.

A particularly simple case is the \textquotedblleft linear
case\textquotedblright\ (Kolotilin et al. (2022)), in which $V(\omega
,a)=V(a)$ and $U_{a}(\omega ,a)=\omega -a.$\footnote{%
More generally, $U_{a}(\omega ,a)$ can be any linear function of $\omega $
and $a$.} This shape of $U_{a}$ arises, for example, in the classical case
of a quadratic loss function of the receiver: $U(\omega ,a)=-\frac{1}{2}%
(a-\omega )^{2}$. Then $a^{\ast }(\mathbb{E}(\omega |m))=\mathbb{E}(\omega
|m)$, and the convexity of $\widehat{V}(\cdot )$ is equivalent to the
convexity of $V(a)$. Following Kolotilin et al. (2022), assume $A=\Omega
=[0,1]$. Hence, the necessary and sufficient condition for the optimality of
full disclosure in the \textquotedblleft linear case\textquotedblright\ can
be written as%
\begin{equation}
V^{\prime }(a_{1})\leq V^{\prime }(a_{2})\text{ for any }a_{1}\in
(0,1),a_{2}\in (0,1)\text{, such that }a_{1}<a_{2}.  \label{iff_linear_case}
\end{equation}%
In this context, our condition (\ref{THE WEAK CONDITION}) becomes

\begin{equation}
\text{For all }a_{1},\ a_{2},\ \omega _{1},\ \omega _{2},\left\{ 
\begin{array}{c}
a_{1}<a_{2} \\ 
\omega _{1}-a_{1}<0<\omega _{2}-a_{2}%
\end{array}%
\right. \Rightarrow V^{\prime }(a_{1})\leq V^{\prime }(a_{2})
\label{our_condition_linear_case}
\end{equation}%
At first sight, (\ref{our_condition_linear_case}) seems weaker than (\ref%
{iff_linear_case}) due to the extra restriction before the implication sign, 
$\omega _{1}-a_{1}<0<\omega _{2}-a_{2}$. Note however that if it were truly
weaker, it would be wrong, because (\ref{iff_linear_case}) is a necessary
condition. But for any $a_{1}\in (0,1),\ a_{2}\in (0,1)$, such that $%
a_{1}<a_{2}$, one can always pick $\omega _{1}$ and $\omega _{2}$ such that $%
\omega _{1}-a_{1}<0<\omega _{2}-a_{2}$.\ Hence, $\omega _{1}-a_{1}<0<\omega
_{2}-a_{2}$ becomes redundant in (\ref{our_condition_linear_case}). The
bottom line is that our sufficient condition for optimality of full
disclosure is in fact necessary and sufficient in the \textquotedblleft
linear case\textquotedblright .

\subsection{\textquotedblleft Linear receiver case\textquotedblright}

Another simple case is what Kolotilin et al. (2022) call the
\textquotedblleft linear receiver case\textquotedblright : $U_{a}(\omega
,a)=\omega -a$ but $V$ may depend on $\omega $. As Kolotilin et al. show, a
sufficient condition for full disclosure to be optimal is that the sender's
utility is convex in $a$ and supermodular in $(a,\omega )$, that is%
\begin{equation}
\left\{ 
\begin{array}{c}
V_{a}(\omega ,a_{1}\dot{)}\leq V_{a}(\omega ,a_{2}\dot{)}\text{ for any }%
\omega \text{ and }a_{1}<a_{2} \\ 
V_{a}(\omega _{1},a\dot{)}\leq V_{a}(\omega _{2},a\dot{)}\text{ for any }a%
\text{ and }\omega _{1}<\omega _{2}%
\end{array}%
\right.  \label{simple receiver Kolotilin}
\end{equation}%
In this context, (\ref{THE WEAK CONDITION}) becomes

\begin{equation}
\text{For all }a_{1},\ a_{2},\ \omega _{1},\ \omega _{2},\left\{ 
\begin{array}{c}
a_{1}<a_{2} \\ 
\omega _{1}-a_{1}<0<\omega _{2}-a_{2}%
\end{array}%
\right. \Rightarrow V_{a}(\omega _{1},a_{1})\leq V_{a}(\omega _{2},a_{2})
\label{condition simple receiver}
\end{equation}%
Our condition is weaker because it requires $V_{a}(\omega ,a)$ to (weakly)
increase only when $a$ grows \emph{and }$\omega $ grows more than $a$, more
precisely from being smaller to being larger than $a$. In particular, our
condition requires neither convexity of $V(\omega ,a)$ in $a$, nor its
supermodularity in $a$ and $\omega $. For example, take the classical
setting of Crawford and Sobel (1982) with $U(\omega ,a)=-(\omega -a)^{2}$
and $V(\omega ,a)=-(\omega -a-b)^{2}$ with $b\geq 0$. These preferences
satisfy the assumptions of the \textquotedblleft simple receiver
case\textquotedblright . The condition from Kolotilin et al. (2022) does not
hold because the sender's utility is concave in $a$. Instead, our condition
is satisfied, as $V_{a}(\omega _{1},a_{1})\leq V_{a}(\omega _{2},a_{2})$
becomes simply $\omega _{1}-a_{1}\leq \omega _{2}-a_{2}$. Although there is
a disagreement between the sender and the receiver regarding the optimal
action in each state, full disclosure is nonetheless optimal, and our
condition sheds light on why it is so: News about the state move the
marginal utilities of the two parties in the same direction, therefore the
decrease of action under \textquotedblleft bad\textquotedblright\ news has a
lower impact on the utility of the sender than the increase of action under
\textquotedblleft good\textquotedblright\ news.

\section{Application to a principal-agent model\label{PA model}}

In this section we explore the implications of our results in the following
principal-agent setting. An agent exerts effort $a$ to produce output $%
y(\omega ,a)$. He bears the cost of effort, which is normalized to be $a$
(in other words, $a$ should be treated as disutility of effort). The agent
receives wage $w(y)$, and the principal receives $y-w(y)$. The agent's and
the principal's utilities of money are (weakly) concave functions $u(\cdot )$
and $v(\cdot )$ respectively. The agent does not know $\omega $, while the
principal does and can send a message to the agent before he chooses effort.
So, the agent is the receiver and the principal is the sender.

For simplicity, we assume that the wage is linear, that is, the agent
receives a fixed share $\delta $ of the output. While we take the
compensation scheme for the agent as given, the conclusions about the
optimality of full disclosure will not depend on $\delta $, as we will see.
However, allowing for a non-linear wage schedule and jointly solving for the
optimal wage schedule and disclosure policy could be an interesting avenue
for future research.

We will first examine the implications of the parties' risk-aversion for the
optimality of full disclosure, given a simple and meaningful production
function. Then we will simplify the parties' preferences by assuming their
risk-neutrality and focus on the properties of the production function
instead.

\subsection{Effects of risk aversion in a simple setting\label{CRRA}}

Consider the following setting:%
\begin{eqnarray*}
y(\omega ,a) &=&\omega a^{\kappa },\ \kappa \in (0,1),\ w(y)=\delta y \\
u(x) &=&\frac{x^{1-\gamma }}{1-\gamma },\ v(x)=\frac{x^{1-\rho }}{1-\rho }
\end{eqnarray*}%
That is, both the agent and the principal exhibit CRRA with coefficients $%
\gamma $ and $\rho $ respectively, where both $\gamma $ and $\rho $ are
non-negative and different from 1. Assume that the upper boundary of $A$ is
large enough to ensure the interior solution of the agent's problem.

We can compute:\ 
\begin{eqnarray*}
U(\omega ,a) &=&\frac{1}{1-\gamma }(\delta \omega )^{1-\gamma }a^{\kappa
(1-\gamma )}-a, \\
U_{a}(\omega ,a) &=&\kappa (\delta \omega )^{1-\gamma }a^{\kappa (1-\gamma
)-1}-1, \\
U_{aa}(\omega ,a) &=&(\kappa (1-\gamma )-1)\kappa (\delta \omega )^{1-\gamma
}a^{\kappa (1-\gamma )-2}, \\
V(\omega ,a) &=&\frac{1}{1-\rho }((1-\delta )\omega )^{1-\rho }a^{\kappa
(1-\rho )}, \\
V_{a}(\omega ,a) &=&\kappa ((1-\delta )\omega )^{1-\rho }a^{\kappa (1-\rho
)-1}.
\end{eqnarray*}%
Notice that the principal's utility cannot be expressed as a function of the
posterior mean, so we cannot use the familiar convexity/non-convexity
argument to establish the optimality/suboptimality of full disclosure.

With some algebra, one can derive%
\begin{equation*}
\frac{V_{a}(\omega ,a)}{-U_{aa}(\omega ,a)}=const\cdot (U_{a}(\omega ,a)+1)^{%
\frac{\gamma -\rho }{1-\gamma }}\cdot a^{\frac{1-\rho }{1-\gamma }},
\end{equation*}%
where $const$ is a positive constant.

It is straightforward to check that the ratio is increasing as both $U_{a}$
and $a$ go up when $\rho \leq \gamma <1$ or $\rho \geq \gamma >1$. Hence, in
this case, (\ref{THE WEAK CONDITION}) holds, and full disclosure is optimal
(see Figure 2). At the same time, under $\rho <1<\gamma $ or $\gamma <1<\rho 
$, the ratio is decreasing when both $U_{a}$ and $a$ increase. According to
Theorem \ref{non-transparency theorem}, full disclosure is then suboptimal.
In all other cases, the ratio is decreasing in $U_{a}$ and increasing in $a$%
. Then, neither (\ref{THE WEAK CONDITION}) nor (\ref{non-transparency}) is
satisfied, and our analysis is inconclusive in such cases.

\FRAME{dtbpFU}{5.7467in}{3.5111in}{0pt}{\Qcb{Figure 2.}}{}{Figure 2}{\special%
{language "Scientific Word";type "GRAPHIC";maintain-aspect-ratio
TRUE;display "USEDEF";valid_file "T";width 5.7467in;height 3.5111in;depth
0pt;original-width 6.4965in;original-height 3.9583in;cropleft "0";croptop
"1";cropright "1";cropbottom "0";tempfilename 'fig2.wmf';tempfile-properties
"XNPR";}}

We can notice that full disclosure fails to be optimal when $\rho $ and $%
\gamma $ are on the opposite sides from $1$. This is related to the fact
that, in this case, state end effort are complements for one party and
substitutes for the other, which can be seen by examining the expressions
for $U_{a}(\omega ,a)$ and $V_{a}(\omega ,a)$. In contrast, when $\rho $ and 
$\gamma $ are both smaller or both greater than $1$, the direction of
interaction between state and effort is the same for both parties, and,
thus, full disclosure gets a chance.

For example, consider a typical textbook situation with a risk neutral
principal ($\rho =0$) and a risk averse agent. If the agent is not too risk
averse ($\gamma <1$), full disclosure is optimal. Since state and effort are
complements for both parties, the principal benefits more from effort
exactly when the agent has higher incentives to exert effort. Instead, when
the agent becomes too risk averse ($\gamma >1$), state and effort become
substitutes for the agent. As a result, good news about productivity \emph{%
depress} effort, while the principal benefits more from effort in higher
states. As a result, full disclosure ceases to be optimal.

When the principal is highly risk averse ($\rho >1$) the story is reversed:
now\emph{\ insufficient} risk aversion of the agent ($\gamma <1$) implies
that full disclosure is suboptimal. This is because now the principal
benefits more from effort under \emph{lower} states, while for the agent
state and effort are complements. One needs to make the agent sufficiently
risk averse ($\gamma >1$) to align the interaction of effort and state
between the two parties, so that full disclosure can be optimal.

What is interesting about the case of a highly risk averse principal is that
full disclosure can be optimal despite lowering the expected effort and can
be harmful despite raising the expected effort. Indeed, one can easily
derive that the disclosure of states in the support of any given message
increases the expected effort under $\gamma <1$ and lowers it under $\gamma
>1$. This observation demonstrates that an increase (decrease) in the
average effort due to disclosure is not sufficient to make full disclosure
optimal (suboptimal), as the direction and strength of the interaction
between state and effort in the principal's payoff matters too.

The role of complementarity/substitutability between the action and the
state can also be observed if one carefully looks at our general condition (%
\ref{THE WEAK CONDITION}). The interaction between the action and the state
for the two parties matters because it affects whether $V_{a}(\omega ,a)$
comoves with $U_{a}(\omega ,a)$ when both $a$ and $U_{a}(\omega ,a)$
increase. Specifically, when action and state are complementary for the
receiver, higher $U_{a}(\omega ,a)$ together with higher $a$ imply higher $%
\omega $, meaning that $\omega _{2}>\omega _{1}$ in (\ref{THE WEAK CONDITION}%
). Then, if action and state are complementary for the sender as well,
higher $\omega $ pushes $V_{a}(\omega ,a)$ upwards for given $a$, thereby
relaxing (\ref{THE WEAK CONDITION}). In contrast, if action and state are
substitutes for the sender, higher $\omega $ pushes $V_{a}(\omega ,a)$
downward for given $a$, thereby tightening (\ref{THE WEAK CONDITION}). By
similar logic, if action and state are substitutes for the receiver, (\ref%
{THE WEAK CONDITION}) is more (less) likely to be satisfied when they are
substitutes (complements) for the sender. A\ word of caution: Although the
fact that action and state are complements (or substitutes) for both parties
helps to satisfy (\ref{THE WEAK CONDITION}), it generally implies neither (%
\ref{THE WEAK CONDITION}), nor that full disclosure is optimal.\footnote{%
For example, if $y=\omega \varphi (a)$ and both parties are risk-neutral,
one can show that the sender's payoff can be represented as a function of
just the posterior mean and then derive that full disclosure is optimal if
and only if $\varphi ^{\prime \prime \prime }(a)\varphi ^{\prime }(a)\geq
(\varphi ^{\prime }(a))^{2}$. Hence, despite complementarity between the
state and the action for both parties, full disclosure may be suboptimal.
See also Mensch (2021) for a discussion on the role of complementarities for
the optimality of full disclosure.}

\subsection{Risk neutral agent and principal, separable production function 
\label{risk neutral separable}}

Sometimes it is more convenient to use the derivatives conditions (\ref%
{derivative conditions 1}) or (\ref{derivative conditions 2}) instead of (%
\ref{THE WEAK CONDITION}). This section illustrates how to apply them in a
simple setting. In the previous subsection, we assumed a simple production
function and played with risk aversion of the parties. Let us now assume
that both parties' utilities are linear in output and examine different
production functions instead. Linearity in output for both parties would
arise, for example, in a setting where both parties are risk neutral and the
wage is linear in output.

The utilities of the agent and the principal under these assumptions are: $%
U(\omega ,a)=\delta y(\omega ,a)-a$ and $V(\omega ,a)=(1-\delta )y(\omega
,a) $, respectively, where $\delta $ is a positive constant.

Suppose\footnote{$y$ may also contain a term \textquotedblleft $\alpha
(\omega )$\textquotedblright\ that only depends on $\omega $, but it would
be irrelevant for both parties' choice problems.}%
\begin{equation}
y(\omega ,a)=\beta (\omega )\varphi (a)+\xi (a),  \label{separable}
\end{equation}%
with $\beta (\cdot )>0,\ \beta ^{\prime }(\cdot )>0,\ \varphi (\cdot )>0,\
\varphi ^{\prime }(\cdot )>0,\ \xi ^{\prime }(\cdot )\geq 0,\ \xi ^{\prime
\prime }(\cdot )+\varphi ^{\prime \prime }(\cdot )<0$, $\xi ^{\prime \prime
}(\cdot )\varphi ^{\prime \prime }(\cdot )\geq 0$ ($\xi ^{\prime \prime
}(\cdot )+\varphi ^{\prime \prime }(\cdot )<0$ ensures strict concavity of $%
y(\omega ,a)$). Assume also $y_{a}(\omega ,a)|_{a=\sup A}<1/\delta $ to
ensure that the agent's choice of $a$ is interior. This output function
could be called \textquotedblleft
multiplicatively-additively\textquotedblright\ separable in state and
effort; we will call it just \textquotedblleft separable\textquotedblright ,
for simplicity. Special cases of this form (such as $\omega \sqrt{a}$
employed in the previous subsection) are commonly used in the literature.%
\footnote{%
It is fair to note that our sufficient condition is not the only way to
check for the optimality of full disclosure in this setting. One can show
that the sender's payoff can eventually be represented as a function of
expected $\beta (\omega )$ and then try to check for the convexity of this
function. However, because the function turns out to be cumbersome, this is
a daunting task, in general. For example, it is hard to use when $\varphi
(\cdot )$ and $\xi (\cdot )$ are arbitrary concave power functions, while
our condition is easy to apply, as we demonstrate below.}

Due to our assumptions on $\beta (\cdot )$ and $\varphi (\cdot )$, state and
effort are complements ($U_{a\omega }>0$). Thus, the relevant condition is (%
\ref{derivative conditions 1}), which becomes:%
\begin{equation}
\text{For each }(\omega ,a)\text{, }\left\{ 
\begin{array}{c}
y_{aa\omega }y_{a}\geq y_{a\omega }y_{aa} \\ 
y_{aaa}y_{a\omega }\geq y_{aa\omega }y_{aa}%
\end{array}%
\right. ,  \label{derivative_output}
\end{equation}%
Using (\ref{separable}), condition (\ref{derivative_output}) can be
rewritten as:%
\begin{equation}
\text{For each }(\omega ,a)\text{, }\left\{ 
\begin{array}{c}
\varphi ^{\prime \prime }(a)\xi ^{\prime }(a)\geq \varphi ^{\prime }(a)\xi
^{\prime \prime }(a) \\ 
\beta (\omega )\left[ \varphi ^{\prime \prime \prime }(a)\varphi ^{\prime
}(a)-(\varphi ^{\prime \prime }(a))^{2}\right] \geq \xi ^{\prime \prime
}(a)\varphi ^{\prime \prime }(a)-\xi ^{\prime \prime \prime }(a)\varphi
^{\prime }(a)%
\end{array}%
\right. .  \label{specific case}
\end{equation}%
Now let us check (\ref{specific case})\ for some specific functional forms
of $\varphi (\cdot )$ and $\xi (\cdot )$. As a first example, assume that
both $\varphi (\cdot )$ and $\xi (\cdot )$ are weakly concave power
functions: $\varphi (a)=ha^{\kappa }$,$\ \xi (a)=la^{\tau }$ with $h>0,\
l>0,\ \kappa \in (0,1],~\tau \in \lbrack 0,1]$, such that $\kappa $ and$\
\tau $ are not both $1$ (to ensure the strict concavity of the output). It
is straightforward to derive that the first inequality boils down to $\kappa
\geq \tau $, and the second inequality always holds. Thus, $\kappa \geq \tau 
$ is a sufficient condition for the optimality of full disclosure.

As another example, consider $\varphi (a)=h\cdot \ln a$ and $\xi (a)=l\cdot
\ln a$. Then the first inequality holds as an equality, and it can be easily
checked that the second one is always satisfied. Hence, full disclosure is
always optimal in such a case.

\section{Conclusion}

In this paper, we have addressed the following question: When is it optimal
for a privately-informed sender to commit to full disclosure of her
information to the receiver? We answer with a sufficient condition that can
be interpreted as a minimal alignment of incentives between the sender and
the receiver.

Several recent papers have derived conditions for the optimality of full
disclosure in terms of the sender's \emph{indirect} utility function,
assuming that it only depends on the posterior mean. Our condition, instead,
speaks directly to the primitive incentives of the parties and does not rely
on any assumption on how the state affects them. For this reason, it can be
easily interpreted and verified in applications.

In a principal-agent setting where the principal is privately informed of a
state that affects the productivity of the agent's effort, the optimal
effort of a risk-averse agent depends on the entire shape of his posterior
belief. As a consequence, given a disclosure policy, the indirect utility
function of the principal does not only depend on the posterior mean, and
the conditions that require this cannot be applied. Our condition, along
with an analogous sufficient condition for \emph{sub}optimality of full
disclosure that we derive, can instead be used to study when full disclosure
is optimal and when it is not, and to interpret the results in light of the
risk aversion of the parties. For instance, we find that full transparency
is optimal under the common modeling assumptions of risk-neutrality of the
principal and risk-aversion of the agent, provided that the agent is not too
risk averse (CRRA with the coefficient of relative risk aversion\ below one).

One interesting question is: In a principal-agent relationship, how does the
optimality of full disclosure depend on the compensation scheme for the
agent? More generally, how to jointly determine the optimal compensation
scheme and disclosure policy? This is an avenue for future research.

\section{Appendix}

\begin{proof}[Proof of Lemma \protect\ref{binary split}]
Consider two states, $\omega _{1}$ and $\omega _{2}$, and a message $m$ with
support $\{\omega _{1},\omega _{2}\}$. Let $\pi _{1}:=\Pr (\omega _{1}|m),\
\pi _{2}:=\Pr (\omega _{2}|m),\ \pi _{2}=1-\pi _{1}$. Let the receiver's
optimal actions in states $\omega _{1},\ \omega _{2}$ and under message $m$
be, respectively, $a_{1}^{\ast },\ a_{2}^{\ast }$, and $a^{\ast }$. Due to
our assumptions on $U(\omega ,a)$, each of $a_{1}^{\ast },\ a_{2}^{\ast }$
and $a^{\ast }$ is unique and determined by the corresponding first-order
condition.

If $a_{1}^{\ast }=a_{2}^{\ast }$, revealing the states is inconsequential.
So, without loss of generality, let $a_{2}^{\ast }>a_{1}^{\ast }$. Then,
from the receiver's first-order condition under $m$ and strict concavity of $%
U_{a}=(\omega ,a)$ in $a$, we get $a_{2}^{\ast }>a^{\ast }>a_{1}^{\ast }$.

The sender (weakly) benefits from disclosing $\omega _{1},\ \omega _{2}$
instead of sending $m$ if and only if%
\begin{equation*}
\pi _{1}V(\omega _{1},a^{\ast })+\pi _{2}V(\omega _{2},a^{\ast })\leq \pi
_{1}V(\omega _{1},a_{1}^{\ast })+\pi _{2}V(\omega _{2},a_{2}^{\ast }),
\end{equation*}%
that is,%
\begin{equation}
\pi _{1}[V(\omega _{1},a^{\ast })-V(\omega _{1},a_{1}^{\ast })]\leq \pi
_{2}[V(\omega _{2},a_{2}^{\ast })-V(\omega _{2},a^{\ast })].  \label{jumps2}
\end{equation}%
Write (\ref{jumps2}) as%
\begin{equation}
\int_{a_{1}^{\ast }}^{a^{\ast }}\pi _{1}V_{a}(\omega _{1},a)da\leq
\int_{a^{\ast }}^{a_{2}^{\ast }}\pi _{2}V_{a}(\omega _{2},a)da.
\label{integrals1}
\end{equation}%
Let $x_{1}(a):=\pi _{1}U_{a}(\omega _{1},a)$ and $x_{2}(a):=-\pi
_{2}U_{a}(\omega _{2},a)$. Due to the first-order conditions for the
receiver under $\omega _{1}$, and $\omega _{2}$, we have: $x_{1}(a_{1}^{\ast
})=x_{2}(a_{2}^{\ast })=0$. In addition, the receiver's first-order
condition under message $m$ yields:%
\begin{eqnarray}
\pi _{1}U_{a}(\omega _{1},a^{\ast })+\pi _{2}U_{a}(\omega _{2},a^{\ast })
&=&0  \label{FOC pooling} \\
&\Rightarrow &x_{1}(a^{\ast })=x_{2}(a^{\ast })=:k<0\text{;}  \notag
\end{eqnarray}%
$k<0$ comes from $U_{a}(\omega _{i},a_{i}^{\ast })=0$, $a_{2}^{\ast
}>a^{\ast }>a_{1}^{\ast }$, and strict concavity of $U_{a}(\omega ,a)$ in $a$%
. Then, given that $dx_{1}:=\pi _{1}U_{aa}(\omega _{1},a)da$ and $%
dx_{2}:=-\pi _{2}U_{aa}(\omega _{2},a)da$, (\ref{integrals1}) is equivalent
to%
\begin{equation}
-\int_{k}^{0}\frac{V_{a}(\omega _{1},a_{1}(x_{1}))}{U_{aa}(\omega
_{1},a_{1}(x_{1}))}dx_{1}\leq \int_{k}^{0}\frac{V_{a}(\omega
_{2},a_{2}(x_{2}))}{-U_{aa}(\omega _{2},a_{2}(x_{2}))}dx_{2},
\label{integrals2}
\end{equation}%
where $a_{i}(x_{i})$ is the value of $a$ derived from the definition of $%
x_{i}$, i.e.,%
\begin{equation*}
a_{1}(x_{1}):=U_{a}^{-1}(\omega _{1},x_{1}/\pi _{1}),\
a_{2}(x_{2}):=U_{a}^{-1}(\omega _{2},-x_{2}/\pi _{2}).
\end{equation*}%
So, if $-V(\omega _{1},a_{1}(x_{1}))/U_{aa}(\omega _{1},a_{1}(x_{1}))\leq
-V_{a}(\omega _{2},a_{2}(x_{2}))/U_{aa}(\omega _{2},a_{2}(x_{2}))$ for any $%
x_{1}=x_{2}\in (k,0)$, then (\ref{integrals2}) (hence, (\ref{jumps2})) is
satisfied.

For any $x_{1}\in (k,0),\ x_{2}\in (k,0)$, we have$\ a_{1}(x_{1})\in
(a_{1}^{\ast },a^{\ast }),\ a_{2}(x_{2})\in (a^{\ast },a_{2}^{\ast })$, that
is, $a_{1}(x_{1})<a_{2}(x_{2})$ and $U_{a}(\omega
_{1},a_{1}(x_{1}))<0<U_{a}(\omega _{2},a_{2}(x_{2}))$. This means that (\ref%
{jumps2}) holds for any $\omega _{1},\ \omega _{2}$, and $\pi _{1}$ if the
following condition is satisfied:%
\begin{equation*}
\text{For all }a_{1},\ a_{2},\ \omega _{1},\ \omega _{2}\text{, }\left\{ 
\begin{array}{c}
a_{1}<a_{2} \\ 
U_{a}(\omega _{1},a_{1})<0<U_{a}(\omega _{2},a_{2})%
\end{array}%
\right. \Rightarrow \frac{V_{a}(\omega _{1},a_{1})}{-U_{aa}(\omega
_{1},a_{1})}\leq \frac{V_{a}(\omega _{2},a_{2})}{-U_{aa}(\omega _{2},a_{2})},
\end{equation*}%
which is condition (\ref{THE WEAK CONDITION}).

To ensure that the sender strictly benefits from the split, we need that (%
\ref{jumps2}) holds as a strict inequality. Clearly, for this, we only need
that $\leq $ turns into $<$ in the above condition.
\end{proof}

\bigskip

\begin{proof}[Proof of Theorem \protect\ref{main theorem 2}]
We first prove that, under (\ref{THE WEAK CONDITION}) with strict instead of
weak inequality, full disclosure is optimal for the sender. We do so by
showing that any message $m^{\ast }$ with non-singleton support $\Omega
^{\ast }$ that pools states that induce different actions is suboptimal.

Let $\pi $ denote the posterior probability distribution conditional on $%
m^{\ast }$. Given a function $f$ of states and actions, given a message $m$
and an action $a$, we let $\widetilde{f}(m,a)$ denote the expected value of $%
f(\omega ,a)$ conditional on $m$. Let $a^{\ast }$ denote the agent's optimal
action upon receiving $m^{\ast }$. It is obtained by solving the first-order
condition $\widetilde{U}_{a}(m^{\ast },a)=0$.

If revealing the states in the support of $m^{\ast }$ can change the
receiver's action, then there exist $\omega _{1}^{\ast },\omega _{2}^{\ast
}\in \Omega ^{\ast }$ such that\footnote{%
If a state $\omega _{1}^{\ast }$ in the support of $m^{\ast }$ induces a
lower action than $a^{\ast }$, then there must also be a state $\omega
_{2}^{\ast }$ in the support of $m^{\ast }$ that induces a higher action for 
$a^{\ast \text{ }}$in order to satisfy the first-order condition after $%
m^{\ast }$, and vice versa.} 
\begin{equation*}
U_{a}(\omega _{1}^{\ast },a^{\ast })<0<U_{a}(\omega _{2}^{\ast },a^{\ast }).
\end{equation*}%
Then, by continuity of $U_{a}(\omega ,a)$ in $\omega $, there exists $%
\varepsilon >0$ such that, for all intervals $\Omega _{1},\Omega _{2}$ of
length smaller than $\varepsilon $ whose interiors contain $\omega
_{1}^{\ast },\omega _{2}^{\ast }$,%
\begin{equation}
\forall \left( \omega _{1},\omega _{2}\right) \in \Omega _{1}\times \Omega
_{2}\text{, \ \ }U_{a}(\omega _{1},a^{\ast })<0<U_{a}(\omega _{2},a^{\ast }).
\label{INTERVALS}
\end{equation}%
Note that, since $\omega _{1}^{\ast },\omega _{2}^{\ast }\in \Omega ^{\ast }$%
, $\pi (\Omega _{1})\pi (\Omega _{2})>0$.

For all intervals $\Omega _{1},\Omega _{2}$ of length smaller than $%
\varepsilon $ whose interiors contain $\omega _{1}^{\ast },\omega _{2}^{\ast
}$, let us decompose $m^{\ast }$ into three messages as follows: $%
m_{1},m_{2} $ with supports contained in $\Omega _{1},\Omega _{2}$, plus a
complementary message $m_{c}$ that induces action $a^{\ast }$ (i.e.,$\ 
\widetilde{U}_{a}(m_{c},a^{\ast })=0$) such that $\pi (m_{1})>0,\ \pi
(m_{2})>0,\ \pi (m_{c})>0$. To be precise, by \textquotedblleft
decomposition\textquotedblright\ we mean that these messages are never sent
in states outside $\Omega ^{\ast }$, and, for each $\omega \in \Omega ^{\ast
}$, conditional on $m^{\ast }$ being drawn, one of the three messages is
sent instead of $m^{\ast }$, so that $\Pr (m_{1}|m^{\ast },\omega )+\Pr
(m_{2}|m^{\ast },\omega )+\Pr (m_{c}|m^{\ast },\omega )=1$. Obviously, $\pi
(m_{1})+\pi (m_{2})+\pi (m_{c})=1$.

Such messages can be constructed because for any decomposition of $m$ into $%
m_{1},m_{2},m_{c}$,%
\begin{equation*}
\widetilde{U}_{a}(m^{\ast },a^{\ast })\equiv \pi (m_{1})\widetilde{U}%
_{a}(m_{1},a^{\ast })+\pi (m_{2})\widetilde{U}_{a}(m_{2},a^{\ast })+\pi
(m_{c})\widetilde{U}_{a}(m_{c},a^{\ast })=0,
\end{equation*}%
and if the supports of $m_{1},m_{2}$ are contained in $\Omega _{1},\Omega
_{2}$, $m_{1}$ and $m_{2}$ satisfy $\widetilde{U}_{a}(m_{1},a^{\ast })<0$
and $\widetilde{U}_{a}(m_{2},a^{\ast })>0$ by (\ref{INTERVALS}). Hence, we
can always adjust $m_{1}$ and $m_{2}$ so that%
\begin{equation}
\pi (m_{1})\widetilde{U}_{a}(m_{1},a^{\ast })+\pi (m_{2})\widetilde{U}%
_{a}(m_{2},a^{\ast })=0\text{ (and hence }\widetilde{U}_{a}(m_{c},a^{\ast
})=0\text{).}  \label{COMP0}
\end{equation}

For every sequence of pairs of intervals $\Omega _{1},\Omega _{2}$ that
contain $\omega _{1}^{\ast }$ and $\omega _{2}^{\ast }$ in their interiors
and have length smaller than $\varepsilon $ and converging to $0$, consider
the corresponding sequence of messages. For each point $(m_{1},m_{2},m_{c})$
of the sequence, consider the relative probabilities%
\begin{equation*}
\frac{\pi (m_{1})}{\pi (m_{1})+\pi (m_{2})},\frac{\pi (m_{2})}{\pi
(m_{1})+\pi (m_{2})}.
\end{equation*}%
The sequence of these probabilities lives in the compact square $\left[ 0,1%
\right] ^{2}$, therefore it has a subsequence that converges to two values $%
p_{1}^{\ast },p_{2}^{\ast }\in \left[ 0,1\right] $ with $p_{1}^{\ast
}+p_{2}^{\ast }=1$. Let $(m_{1}^{n},m_{2}^{n},m_{c}^{n})_{n>0}$ denote the
corresponding subsequence of messages. For each $n>0$, recall from (\ref%
{COMP0}) that%
\begin{equation*}
-\frac{\pi (m_{1}^{n})}{\pi (m_{1}^{n})+\pi (m_{2}^{n})}\widetilde{U}%
_{a}(m_{1}^{n},a^{\ast })=\frac{\pi (m_{2}^{n})}{\pi (m_{1}^{n})+\pi
(m_{2}^{n})}\widetilde{U}_{a}(m_{2}^{n},a^{\ast }).
\end{equation*}%
By continuity of $U_{a}(\omega ,a^{\ast })$ in $\omega $, we have $%
\lim_{n\rightarrow \infty }\widetilde{U}_{a}(m_{i}^{n},a^{\ast
})=U_{a}(\omega _{i}^{\ast },a^{\ast })$ for each $i=1,2$. Hence, we get%
\begin{equation}
-p_{1}^{\ast }U_{a}(\omega _{1}^{\ast },a^{\ast })=p_{2}^{\ast }U_{a}(\omega
_{2}^{\ast },a^{\ast }).  \label{COMPENSATION}
\end{equation}%
Note that this also implies $p_{1}^{\ast },p_{2}^{\ast }\not=0$.

For each $n>0$, call $a_{1}^{n},a_{2}^{n}$ the receiver's optimal actions
under $m_{1}^{n},m_{2}^{n}$, i.e., $\widetilde{U_{a}}(m_{1}^{n},a_{1}^{n})=%
\widetilde{U_{a}}(m_{2}^{n},a_{2}^{n})=0$. Note that $a_{1}^{n}<a^{\ast
}<a_{2}^{n}$, as $\widetilde{U_{a}}(m_{1}^{n},a^{\ast })<0<\widetilde{U_{a}}%
(m_{2}^{n},a^{\ast })$ and $U$ is strictly concave in $a$. The sender's
expected utility increases after the decomposition of $m^{\ast }$ into $%
m_{1}^{n},m_{2}^{n},m_{c}^{n}$ if the following inequality holds:%
\begin{eqnarray*}
\widetilde{V}(m^{\ast },a^{\ast }) &=&\pi (m_{1}^{n})\widetilde{V}%
(m_{1}^{n},a^{\ast })+\pi (m_{2}^{n})\widetilde{V}(m_{2}^{n},a^{\ast })+\Pr
(m^{n})\widetilde{V}(m_{c}^{n},a^{\ast }) \\
&<&\pi (m_{1}^{n})\widetilde{V}(m_{1}^{n},a_{1}^{n})+\pi (m_{2}^{n})%
\widetilde{V}(m_{2}^{n},a_{2}^{n})+\Pr (m^{n})\widetilde{V}%
(m_{c}^{n},a^{\ast }).
\end{eqnarray*}%
Rewrite the inequality as%
\begin{equation*}
\pi (m_{1}^{n})\left[ \widetilde{V}(m_{1}^{n},a^{\ast })-\widetilde{V}%
(m_{1}^{n},a_{1}^{n})\right] <\pi (m_{2}^{n})\left[ \widetilde{V}%
(m_{2}^{n},a_{2}^{n})-\widetilde{V}(m_{2}^{n},a^{\ast })\right] ,
\end{equation*}%
and then as%
\begin{equation}
\pi (m_{1}^{n})\int_{a_{1}^{n}}^{a^{\ast }}\widetilde{V}_{a}(m_{1}^{n},a)da<%
\pi (m_{2}^{n})\int_{a^{\ast }}^{a_{2}^{n}}\widetilde{V}_{a}(m_{2}^{n},a)da.
\label{TO SHOW}
\end{equation}

Call $a_{1}^{\ast },a_{2}^{\ast }$ the receiver's optimal actions under $%
\omega _{1}^{\ast }$ and $\omega _{2}^{\ast }$. For each $i=2$, by
continuity of $U_{a}$ in $\omega $, we have $\lim_{n\rightarrow \infty
}a_{i}^{n}=a_{i}^{\ast }$, and by continuity of $V_{a}$ in $\omega $, we
have $\lim_{n\rightarrow \infty }\widetilde{V}_{a}(m_{i}^{n},a)=V_{a}(\omega
_{i}^{\ast },a)$. Therefore,%
\begin{eqnarray*}
\lim_{n\rightarrow \infty }\pi (m_{1}^{n})\int_{a_{1}^{n}}^{a^{\ast }}%
\widetilde{V}_{a}(m_{1}^{n},a)da &=&p_{1}^{\ast }\int_{a_{1}^{n}}^{a^{\ast
}}V_{a}(\omega _{1}^{\ast },a)da, \\
\lim_{n\rightarrow \infty }\pi (m_{2}^{n})\int_{a^{\ast }}^{a_{2}^{n}}%
\widetilde{V}_{a}(m_{2}^{n},a)da &=&p_{2}^{\ast }\int_{a^{\ast
}}^{a_{2}^{n}}V_{a}(\omega _{2}^{\ast },a)da.
\end{eqnarray*}%
So, it is enough to show%
\begin{equation}
p_{1}^{\ast }\int_{a_{1}^{\ast }}^{a^{\ast }}V_{a}(\omega _{1}^{\ast
},a)da<p_{2}^{\ast }\int_{a^{\ast }}^{a_{2}^{\ast }}V_{a}(\omega _{2}^{\ast
},a)da;  \label{integrals3}
\end{equation}%
then, for sufficiently large $n$, the decomposition satisfies (\ref{TO SHOW}%
).

Thus, we have reduced the problem to checking if decomposing a hypothetical
message with binary support $\{\omega _{1}^{\ast },\omega _{2}^{\ast }\}$
and relative probabilities $p_{1}^{\ast }$ and $p_{2}^{\ast }=1-p_{1}^{\ast
} $ of the two states strictly benefits the sender. To see it, notice that,
if we replace $p_{i}^{\ast }$ with $\pi _{i}$, and $\omega _{i}^{\ast }$
with $\omega _{i}$, (\ref{integrals3}) and (\ref{COMPENSATION}) become (\ref%
{integrals1}) and (\ref{FOC pooling}) from the proof of Lemma \ref{binary
split}, except that (\ref{integrals3}) is a strict inequality while (\ref%
{integrals1}) is a weak inequality. Hence, we arrive at the same sufficient
condition as Lemma \ref{binary split} delivers, except that (as noted at the
end of the proof of the lemma) the inequality between $\dfrac{V_{a}(\omega
_{1},a_{1})}{-U_{aa}(\omega _{1},a_{1})}$ and $\dfrac{V_{a}(\omega
_{2},a_{2})}{-U_{aa}(\omega _{2},a_{2})}$ becomes strict.

The last step of the proof is showing that, if full disclosure is optimal
when condition (\ref{THE WEAK CONDITION}) holds with strict inequality
between the ratios, so it is when it holds with weak inequality. Suppose by
contradiction that full disclosure is suboptimal and condition (\ref{THE
WEAK CONDITION}) holds. Let $\Delta $ denote the difference between the
expected utility of the sender under the optimal communication scheme and
under full disclosure. Fix $\gamma \in (0,1)$ and let%
\begin{equation*}
\hat{V}(\omega ,a)=V(\omega ,a)-\gamma \exp \left( U_{a}(\omega ,a)\right) .
\end{equation*}%
So we have%
\begin{eqnarray*}
\hat{V}_{a}(\omega ,a) &=&V_{a}(\omega ,a)-\gamma U_{aa}(\omega ,a)\exp
\left( U_{a}(\omega ,a)\right) , \\
\frac{\hat{V}_{a}(\omega ,a)}{-U_{aa}(\omega ,a)} &=&\frac{V_{a}(\omega ,a)}{%
-U_{aa}(\omega ,a)}+\gamma \exp \left( U_{a}(\omega ,a)\right) .
\end{eqnarray*}%
Since $\exp \left( U_{a}(\omega ,a)\right) $ is strictly increasing in $%
U_{a}(\omega ,a)$, the following holds:%
\begin{gather*}
\text{For all }a_{1},\ a_{2},\ \omega _{1},\ \omega _{2}\text{ such that }%
U_{a}(\omega _{1},a_{1})<U_{a}(\omega _{2},a_{2})\text{, } \\
\frac{V_{a}(\omega _{1},a_{1})}{-U_{aa}(\omega _{1},a_{1})}\leq \frac{%
V_{a}(\omega _{2},a_{2})}{-U_{aa}(\omega _{2},a_{2})}\Rightarrow \frac{\hat{V%
}_{a}(\omega _{1},a_{1})}{-U_{aa}(\omega _{1},a_{1})}<\frac{\hat{V}%
_{a}(\omega _{2},a_{2})}{-U_{aa}(\omega _{2},a_{2})}.
\end{gather*}%
Therefore, if condition (\ref{THE WEAK CONDITION}) holds with $V$, it holds
with strict inequality with $\hat{V}$. For sufficiently small $\gamma $, the
expected utility of the sender with $V$ and $\hat{V}$ differ in absolute
value by less than $\Delta /2$ no matter the communication scheme, and hence
full disclosure remains suboptimal with $\hat{V}$. But we have shown above
that full disclosure is optimal when condition (\ref{THE WEAK CONDITION})
holds with strict inequality, a contradiction.
\end{proof}

\bigskip

\begin{proof}[Proof of Lemma \protect\ref{derivative lemma}]
Let $h(\omega ,a):=\dfrac{V_{a}(\omega ,a)}{-U_{aa}(\omega ,a)}$. Condition (%
\ref{THE DERIVABLE CONDITION}) is equivalent to the statement that $h(\omega
,a)$ weakly increases in all directions in the $\Omega \times A$ space in
which $a$ and $U_{a}(\omega ,a)$ jointly increase. So, let us define a
direction through a function $\omega (a)$ and take the full derivative of $%
h(\omega (a),a)$ with respect to $a$:%
\begin{equation*}
\frac{dh}{da}=\frac{-\dfrac{dV_{a}}{da}U_{aa}+\dfrac{dU_{aa}}{da}V_{a}}{%
(U_{aa})^{2}}.
\end{equation*}%
We want to show that $\frac{dh}{da}\geq 0$, which is equivalent to%
\begin{equation}
\dfrac{dU_{aa}}{da}V_{a}-\dfrac{dV_{a}}{da}U_{aa}\geq 0,
\label{Derivatives1}
\end{equation}%
for \emph{all} $\omega (a)$ such that $\dfrac{dU_{a}}{da}>0$, i.e., all
directions in which $U_{a}$ increases as well. As $\dfrac{dU_{a}}{da}%
=U_{aa}+U_{a\omega }\dfrac{d\omega }{da}$, we have that $\dfrac{dU_{a}}{da}%
>0 $ is equivalent to%
\begin{equation}
\left\{ 
\begin{array}{c}
\dfrac{d\omega }{da}>-\dfrac{U_{aa}}{U_{a\omega }}\text{ if }U_{a\omega }>0
\\ 
\dfrac{d\omega }{da}<-\dfrac{U_{aa}}{U_{a\omega }}\text{ if }U_{a\omega }<0%
\end{array}%
\right.  \label{Bifurcation}
\end{equation}%
If $U_{a\omega }=0$, $\dfrac{dU_{a}}{da}$ cannot be positive, as $U_{aa}<0$
by assumption.

Taking into account that%
\begin{eqnarray*}
\dfrac{dV_{a}}{da} &=&V_{aa}+V_{a\omega }\frac{d\omega }{da} \\
\dfrac{dU_{aa}}{da} &=&U_{aaa}+U_{aa\omega }\frac{d\omega }{da}
\end{eqnarray*}%
inequality (\ref{Derivatives1}) becomes%
\begin{eqnarray}
&&U_{aaa}V_{a}+U_{aa\omega }V_{a}\frac{d\omega }{da}-\left(
V_{aa}U_{aa}+V_{a\omega }U_{aa}\frac{d\omega }{da}\right)  \notag \\
&\equiv &U_{aaa}V_{a}-V_{aa}U_{aa}+(U_{aa\omega }V_{a}-V_{a\omega }U_{aa})%
\frac{d\omega }{da}\geq 0.  \label{general new}
\end{eqnarray}

Consider first the case when $U_{a\omega }>0$. Then, the necessary and
sufficient conditions for (\ref{general new}) to hold for \emph{all} $\omega
(a)$ such that $\dfrac{dU_{a}}{da}>0$, given that by (\ref{Bifurcation}) $%
\dfrac{d\omega }{da}$ can take \emph{all} values above $-\dfrac{U_{aa}}{%
U_{a\omega }}$, are the following:%
\begin{equation*}
\left\{ 
\begin{array}{c}
U_{aa\omega }V_{a}-V_{a\omega }U_{aa}\geq 0 \\ 
U_{aaa}V_{a}-V_{aa}U_{aa}-(U_{aa\omega }V_{a}-V_{a\omega }U_{aa})\dfrac{%
U_{aa}}{U_{a\omega }}\geq 0%
\end{array}%
\right. ,
\end{equation*}%
which becomes%
\begin{equation}
\left\{ 
\begin{array}{c}
U_{aa\omega }V_{a}\geq V_{a\omega }U_{aa} \\ 
V_{a}(U_{aaa}U_{a\omega }-U_{aa\omega }U_{aa})\geq U_{aa}(V_{aa}U_{a\omega
}-V_{a\omega }U_{aa})%
\end{array}%
\right. .  \label{complements}
\end{equation}

Consider now the case when $U_{a\omega }<0$. Then, following the same steps
we get%
\begin{equation*}
\left\{ 
\begin{array}{c}
U_{aa\omega }V_{a}\leq V_{a\omega }U_{aa} \\ 
V_{a}(U_{aaa}U_{a\omega }-U_{aa\omega }U_{aa})\leq U_{aa}(V_{aa}U_{a\omega
}-V_{a\omega }U_{aa})%
\end{array}%
\right. ,
\end{equation*}
\end{proof}

\bigskip

\begin{proof}[Proof of Theorem \protect\ref{non-transparency theorem}]
Suppose that there exists a pair of states $\omega _{1},\omega _{2}$ with $%
a_{1}^{\ast }(\omega _{1})<a_{2}^{\ast }(\omega _{2})$ such that, for all $%
a_{1},\ a_{2}$ satisfying $a_{1}<a_{2}$ and $U_{a}(\omega
_{1},a_{1})<0<U_{a}(\omega _{2},a_{2})$,%
\begin{equation*}
\frac{V_{a}(\omega _{1},a_{1})}{-U_{aa}(\omega _{1},a_{1})}>\frac{%
V_{a}(\omega _{2},a_{2})}{-U_{aa}(\omega _{2},a_{2})}\text{.}
\end{equation*}%
For such states, (\ref{integrals2}) does not hold, and hence (\ref{jumps2})
does not hold, which means that pooling those states is better than
revealing them. Then, by continuity, pooling intervals whose interiors
contain (respectively) $\omega _{1}$ and $\omega _{2}$ is better than
revealing the states in the intervals. Since $\omega _{1}$ and $\omega _{2}$
are in the support of the prior, such intervals have positive measure, and
thus full disclosure is suboptimal.
\end{proof}

\end{document}